\documentclass[a4paper,12pt]{article}
\pdfoutput=1
\usepackage{geometry}
\usepackage{amsmath,amssymb,amsfonts}
\usepackage{xcolor,graphicx,cite,soul}
\usepackage{array,booktabs,longtable}
\usepackage{caption,microtype}
\usepackage{hyperref}
\usepackage{axodraw}

\newcommand\tb{\tan\beta}

\newcommand\ReDiag{\mathop{%
  \raise .5pt\hbox{[}%
  \widetilde{\mathrm{Re}}%
  \raise .5pt\hbox{]}}}
\newcommand\ReOffDiag{\mathop{%
  \raise .5pt\hbox{$\llbracket$}%
  \widetilde{\mathrm{Re}}%
  \raise .5pt\hbox{$\rrbracket$}}}

\newcommand\MA{M_A}

\newcommand\stau{\tilde \tau}

\renewcommand\stop{\tilde t}

\newcommand\ino[1]{\tilde\chi_{#1}}

\newcommand\chapm[1]{\ino{#1}^\pm}

\newcommand\cha{\chapm}

\newcommand\neu[1]{\ino{#1}^0}
\newcommand\mneu[1]{m_{\neu{#1}}}

\newcommand\refta[1]{Tab.~\ref{#1}}
\newcommand\reftas[1]{Tabs.~\ref{#1}}
\newcommand\refse[1]{Sect.~\ref{#1}}

\newcommand\citere[1]{Ref.~\cite{#1}}
\newcommand\citeres[1]{Refs.~\cite{#1}}

\newcommand\eg{e.g.}

\newcommand{\CP}{{\cal CP}}
\newcommand{\cp}{{\CP}}

\newcommand{\tev}{\,\, \mathrm{TeV}}
\newcommand{\gev}{\,\, \mathrm{GeV}}

\newcommand\Sq{{\tilde q}}
\newcommand\msq[1]{m_{\Sq_{#1}}}

\newcommand{\br}{\text{BR}}

\newcommand{\sig}{\sigma}

\def\reffi#1{\mbox{Fig.~\ref{#1}}}

\definecolor{Orange}{named}{orange}
\definecolor{Purple}{named}{purple}
\definecolor{Lightblue}{cmyk}{0.9,0.1,0.1,0.3}
\definecolor{dgelborange}{cmyk}{0.,0.3,0.5, 0.}
\definecolor{Lila}{rgb}{0.5,0.,1}

\graphicspath{{figs/}}
\renewcommand{\arraystretch}{1.2}

\newcommand{\mgrav}{\ensuremath{m_{3/2}}}
\newcommand{\Min}{\ensuremath{M_{\rm in}}}
\newcommand{\MGUT}{\ensuremath{M_{\rm GUT}}}
\newcommand{\ETslash}{\ensuremath{/ \hspace{-.7em} E_T}}
\newcommand{\msqt}{\ensuremath{m_{\tilde q_3}}}

\newcommand{\mslep}{\ensuremath{m_{\tilde \ell}}}
\newcommand{\msmu}[1]{\ensuremath{m_{\tilde \mu_{#1}}}}
\newcommand{\msel}[1]{\ensuremath{m_{\tilde e_{#1}}}}
\newcommand{\msusy}{M_{\rm SUSY}}
\newcommand{\mst}[1]{m_{\tilde t_{#1}}}

\newcommand{\gmt}{\ensuremath{(g-2)_\mu}}

\newcommand{\bmm}{\ensuremath{\br(B_s \to \mu^+\mu^-)}}

\newcommand{\ssi}{\ensuremath{\sigma^{\rm SI}_p}}

\newcommand{\ssd}{\ensuremath{\sigma^{\rm SD}_p}}

\allowdisplaybreaks
\begin{document}
\thispagestyle{empty}

\def\thefootnote{\fnsymbol{footnote}}

\begin{flushright}
\mbox{}
IFT--UAM/CSIC--18-003 \\
\end{flushright}

\vspace{0.5cm}

\begin{center}

{\large\sc 
  {\bf New SUSY Fits for the ILC and CLIC}}



\vspace{1cm}

{\sc
S.~Heinemeyer$^{1\,2\,3\,}$%
\footnote{email: Sven.Heinemeyer@cern.ch}%
\footnote{Talk presented on behalf of the {\tt MasterCode} collaboration
at the International Workshop\\
\mbox{}\hspace{5mm} on Future Linear
Colliders (LCWS2017), Strasbourg, France, 23-27 October 2017.}
}

\vspace*{.7cm}

{\sl
$^1$Campus of International Excellence UAM+CSIC, 
Cantoblanco, 28049, Madrid, Spain 

\vspace*{0.1cm}

$^2$Instituto de F\'isica Te\'orica (UAM/CSIC), 
Universidad Aut\'onoma de Madrid, \\ 
Cantoblanco, 28049, Madrid, Spain

\vspace*{0.1cm}

$^3$Instituto de F\'isica de Cantabria (CSIC-UC), 
39005, Santander, Spain


}

\end{center}

\vspace*{0.1cm}

\begin{abstract}
\noindent
We review the {\tt MasterCode} fits of several incarnations of the Minimal
Supersymmetric Standard Model (MSSM). These include the GUT models based on
mAMSB and SU(5), sub-GUT models as well as a model defined at low energies
with 11 free parameters, the pMSSM11. The fit combines consistently
measurements of Higgs boson properties, searches for additional Higgs bosons
and supersymmetric (SUSY) particles, low-energy and flavor experiments as well
as Dark Matter (DM) measurements. We predict the preferred SUSY mass spectra
in these models and analyze the discovery potential of future $e^+e^-$
colliders such as the ILC and CLIC.  
\end{abstract}


\def\thefootnote{\arabic{footnote}}
\setcounter{page}{0}
\setcounter{footnote}{0}

\newpage


\section{Introduction}
\label{sec:intro}

Models invoking the appearance of supersymmetry (SUSY) at the TeV scale
are being tested by the so far negative results of high-sensitivity searches
for sparticles at the LHC~\cite{CMSWiki,ATLASWiki} and for the
scattering of dark matter particles~\cite{lux16,PICO,XENON1T,pandax}.
There have been many global analyses of the implications of these experiments
for specific SUSY models, mainly within the minimal supersymmetric
extension of the Standard Model (MSSM), in which the lightest supersymmetric
particle (LSP) is stable and a candidate for cold dark matter
(CDM). This may well be the lightest neutralino, $\neu1$~\cite{EHNOS},
as is assumed here. 
Some of these studies have assumed universality of the soft SUSY-breaking
parameters at the GUT scale, e.g., in the constrained MSSM (the
CMSSM)~\cite{mcold,mc9,Fittino,GAMBIT} and in models with non-universal Higgs
masses (the NUHM1,2)~\cite{mc9,mc10}. Other analyses have taken a
phenomenological approach, allowing free variation in the soft SUSY-breaking
parameters at the electroweak scale (the
pMSSM)~{\cite{pMSSM,mc11,gambit-pmssm7}}. Here we review the results of
studies of the GUT based minimal scenario for anomaly-mediated SUSY breaking
(the mAMSB)~\cite{mAMSB}, of the GUT based SU(5) framework~\cite{SU5}, 
of SUSY models, in which universality of the
soft SUSY-breaking parameters is imposed at some input scale \Min\ below
the GUT scale \MGUT\ but above the electroweak scale~\cite{sub-GUT,ELOS},
which are termed ``sub-GUT'' models, as well as of a ``phenomenological
MSSM''~\cite{pMSSM} defined at low energies with 11 independent parameters, the
pMSSM11. These studies have been presented in \citeres{mc13,mc14,mc16,mc15},
respectively. 

These studies have been performed using the {\tt MasterCode} 
framework~\cite{mcold,mc9,mc10,mc11,mc12,mc13,mc14,mc15,mc16,mcweb}. 
{\tt MasterCode} combines consistently experimental data from LHC Higgs-boson
measurements, searches for additional Higgs bosons and SUSY
particles at the LHC, low-energy and flavor experiments as well 
as CDM measurements. 
In all models we predict the preferred SUSY
mass spectra at the $1$~and $2\,\sig$ level~\cite{mc13,mc14,mc16,mc15}.

Assuming that SUSY is realized in nature and the scalar quarks and/or the
gluino are in the kinematic reach of the LHC, it is expected that
these strongly interacting particles are eventually produced and studied.
On the other hand, SUSY particles that interact only via the electroweak
force, \eg, the scalar leptons or charginos/neutralinos, have a much
smaller 
production cross section at the LHC.  Correspondingly, the LHC
discovery potential as well as the current experimental bounds are
substantially weaker~\cite{CMSWiki,ATLASWiki}. 
However, at a (future) $e^+e^-$ collider , depending on their masses 
and the available center-of-mass energy, the electroweak SUSY particles
could be produced and 
analyzed in detail.  Corresponding studies can be found for the ILC 
in \citeres{ILC-TDR,teslatdr,Ac2004,ilc1,ilc2,LCreport} and for CLIC 
in \citeres{CLIC1,CLIC2,LCreport}. Such precision studies will be 
crucial to determine their nature and the underlying (SUSY) parameters.
Our analyses show which SUSY particles or additional Higgs bosons might
be in the reach of the ILC or CLIC, depending on the assumed scenario.

It should be kept in mind that a precision analysis at ILC or CLIC also
requires predictions of the production and decay properties of the SUSY
particles or additional Higgs bosons with high precision. At least full
one-loop calculations are necessary. A set of corresponding
{\em complete and consistent} one-loop calculations (allowing also for
complex parameters) 
has been published over the last years for SUSY/Higgs production cross
sections at $e^+e^-$ colliders~\cite{HS-Prod} and SUSY/Higgs
decays~\cite{HS-Decay} (based on \citere{MSSMCT}).


\section{The models under investigation}
\label{sec:models}

All our models are certain realizations of the general
MSSM~\cite{Ni1984,Ba1988,HaK85,GuH86}, which predicts two scalar partners for
all SM fermions as well as fermionic partners to all SM bosons. 
In particular two scalar quarks or scalar leptons are predicted for each
SM quark or lepton. Concerning the Higgs-boson sector, 
contrary to the case of the SM, in the MSSM two Higgs doublets are required.
This results in five physical Higgs bosons instead of the single Higgs
boson in the SM.  These are the light and heavy $\cp$-even Higgs bosons, 
$h$ and $H$, the $\cp$-odd Higgs boson, $A$, and the charged Higgs bosons,
$H^\pm$.
The Higgs sector of the MSSM is described at the tree level by two
parameters: 
the mass of the $\cp$-odd Higgs boson, $\MA$, and the ratio of the two
vacuum expectation values, $\tb = v_2/v_1$. Higher-order corrections are
crucial to yield reliable predictions in the MSSM Higgs-boson sector, see
\citeres{habilSH,awb2,PomssmRep} for reviews.
The lightest Higgs boson, $h$ can be identified~\cite{Mh125} with 
the particle  discovered at the LHC~\cite{ATLASdiscovery,CMSdiscovery} 
with a mass around $\sim 125\gev$~\cite{MH125}.
The neutral SUSY partners of the (neutral) Higgs and electroweak gauge
bosons are the four neutralinos, $\neu{1,2,3,4}$.  The corresponding
charged SUSY partners are the charginos, $\cha{1,2}$. In the results
reviewed here the Higgs mixing parameter $\mu$ is assumed to be positive.
Furthermore, in all our models 
we assume the Minimal Flavor Violation (MFV) scenario in which
generation mixing is described by the Cabibbo-Kobayashi-Maskawa (CKM) matrix.
This is motivated by phenomenological constraints on low-energy
flavor-changing neutral interactions.


\subsection{The mAMSB}
\label{sec:mAMSB}

In the mAMSB there are 3 relevant continuous parameters, the gravitino mass,
\mgrav, which sets the scale of SUSY breaking, the supposedly universal soft
SUSY-breaking scalar mass, $m_0$, and the ratio of Higgs vacuum expectation
values, $\tb$. The sampled parameter ranges are shown in
\refta{tab:mamsb-ranges}. 

The LSP is either a Higgsino-like or a wino-like neutralino $\neu1$.
In both cases the $\neu1$ is almost mass degenerate with its chargino partner,
$\cha1$. 
Within this mAMSB framework, it is well known that if one requires that
a wino-like $\neu1$ is the dominant source of the CDM density indicated by
Planck measurements of the cosmic microwave background radiation, namely
$\Omega_{\rm CDM} h^2 = 0.1186 \pm 0.0020$~\cite{Planck15}, $\mneu1 \simeq 3
\tev$~\cite{winomass1,winomass} after inclusion of Sommerfeld enhancement
effects~\cite{Sommerfeld1931}. If instead the CDM density is to be explained
by a Higgsino-like $\neu1$, $\mneu1$ takes a value of about $1.1 \tev$.  

\begin{table}[htb!]
\begin{center}
\begin{tabular}{|c|c|c|} \hline
Parameter   &  \; \, Range      	& Segments \\
\hline         
$m_0$       &  ( 0.1  , $50 \tev$)    & 4 \\
$m_{3/2}$   &  ( 10  , $1500 \tev$) & 3 \\
$\tb$         &  ( 1  , 50)      & 4 	\\
\hline \hline
\multicolumn{2}{|c|}{Total number of boxes}    	& 48 \\
\hline
\end{tabular}
\caption{\it Ranges of the mAMSB parameters sampled, together with the
  numbers of segments into which each range was divided (see Sect.~3),
  and the corresponding number of sample boxes.} 
\label{tab:mamsb-ranges}
\end{center}
\end{table}


\subsection{The GUT based SU(5)}
\label{sec:SU5}

Here we assume a universal, SU(5)-invariant gaugino mass parameter $m_{1/2}$,
which is input at the GUT scale, as are the other SUSY-breaking
parameters:
we assume the conventional multiplet assignments of matter fields in the
minimal superymmetric GUT: 
\begin{equation}
(q_L, u^c_L, e^c_L)_i \; \in \; \mathbf{10}_i, \; \; 
(\ell_L, d^c_L)_i \; \in \; \mathbf{\bar 5}_i \, ,
\label{assignments}
\end{equation}
where the subscript $i = 1, 2, 3$ is a generation index. The only
relevant Yukawa couplings are those of the third generation,
particularly that of the $t$ quark (and possibly the $b$ quark 
and the $\tau$ lepton) that may play an important role
in generating electroweak symmetry breaking. 
Following the MFV scenario, we assume 
that the soft SUSY-breaking scalar masses for
the different $\mathbf{10}_i$ and $\mathbf{\bar 5}_i$ representations are
universal in generation space, and are denoted by
$m_{10}$ and $m_{5}$, respectively. In contrast to the CMSSM, NUHM1 and
NUHM2, we allow $m_5 \ne m_{10}$. We assume a universal soft trilinear
SUSY-breaking parameter~$A_0$. 

We assume the existence of two Higgs doublets $H_u$ and $H_d$ in
$\mathbf{5}$ and $\mathbf{\bar 5}$ representations that break
electroweak symmetry and give masses to the charge +2/3 and charge -1/3
and -1 matter fields, respectively. In the absence of any
phenomenological constraints, we allow the soft SUSY-breaking 
contributions to the $H_u$ and $H_d$ masses, $m_{H_u}$ and $m_{H_d}$, to
be different from each other, as in the NUHM2, as well as from $m_5$ and
$m_{10}$. As in the CMSSM, NUHM1 and NUHM2, we allow the ratio of Higgs
vacuum expectation values, $\tan \beta$, to be a free parameter.
The sampled parameter ranges are shown in \refta{tab:su5-ranges}.

\begin{table}[htb!]
\begin{center}
\begin{tabular}{|c|c|c|} 
\hline
Parameter   &  Range      & Segments  \\ 
\hline         
$m_{1/2}$        &  $( 0  , 4) \tev$     & {2} \\
$m_{5}$        &  $( - 2.6  , 8) \tev$     & 2 \\
$m_{10}$        &  $( - 1.3 , 4) \tev$     & 3 \\
$m_{H_u}$        &  $( -7  , 7) \tev$     & 3 \\
$m_{H_d}$        &  $( -7  , 7) \tev$     & 3 \\
$A_0$        	&  $( -8  , 8) \tev$      & 1 \\
$\tb$         &  $( 2  , 68)$      & 1 \\
\hline \hline
\multicolumn{2}{|c|}{Total number of boxes} & 108 \\
\hline
\end{tabular}
\caption{\it Ranges of the SUSY SU(5) GUT parameters sampled, together
  with the numbers of segments into which each range was divided, and
  the corresponding total number of sample boxes.}
\label{tab:su5-ranges}
\end{center}
\end{table}


\subsection{The sub-GUT model}
\label{sec:subGUT}

This class of models is based on the CMSSM, but the universality of the
soft SUSY-breaking parameters is imposed at some input scale \Min\ below
the GUT scale \MGUT\ but above the electroweak scale~\cite{sub-GUT,ELOS}.
The sampled parameter ranges are shown in \refta{tab:subGUT-ranges}.

This type of models is 
well motivated theoretically, since the soft SUSY-breaking
parameters in the visible sector may be induced by some dynamical mechanism
such as gluino condensation that kicks in below the GUT scale. Specific
examples of sub-GUT models include warped extra dimensions~\cite{wed}
and mirage mediation~\cite{mirage}. 

Sub-GUT models are of particular phenomenological interest, since the reduction
in the amount of renormalization-group (RG) running below \Min, compared to
that below \MGUT\ in the CMSSM and related models, leads naturally
to SUSY spectra that are more compressed~\cite{sub-GUT}. These may offer
extended possibilities for `hiding' SUSY via suppressed
\ETslash\ signatures, as well as offering enhanced possibilities for
different coannihilation processes.
 
\begin{table}[htb!]
\begin{center}
\begin{tabular}{|c|c|c|}
\hline
Parameter   &  \; \, Range      & Segments  \\
\hline 
	\Min      &  $(10^{3}, 10^{16} ) \gev$  & 6 \\
$m_{1/2}$       &  $(0, 6) \tev$  & 2 \\
$m_0$       &  $(0, 6) \tev$  & 2 \\
$A_0$        &  $(- 15,  10) \tev$  & 2 \\
$\tb$         &  ( 1  , 60)      & 2 \\
\hline\hline
\multicolumn{2}{|c|}{Total number of boxes}& 96 \\
\hline
\end{tabular}
\caption{\it The ranges of the sub-GUT MSSM parameters sampled, together
  with the numbers of segments into which they are divided, together
  with the total number of sample boxes shown in the last row. 
}
\label{tab:subGUT-ranges}
\end{center}
\end{table}


\subsection{The pMSSM11}
\label{sec:pMSSM11}

As mentioned above, in this paper we consider a pMSSM scenario with
eleven independent parameters, namely
\begin{align}
{\rm 3~gaugino~masses}&: \; M_{1,2,3} \, ,  \nonumber \\
{\rm 2~squark~masses}&: \; {\msq \, \equiv} \, m_{\tilde q_1}, m_{\tilde q_2} \nonumber \\
                       &\ne \, \msqt \, = \, m_{\tilde t} , m_{\tilde b}, \nonumber \\
{\rm 2~slepton~masses}&: \; {\mslep} \, \equiv \, m_{\tilde \ell_1} \, = m_{\tilde \ell_2} \, = \, \msel, \msmu
\nonumber \\
                       &\ne \, m_{\ell_3} \, = \, m_{\tilde \tau}, \nonumber \\
\label{mc11}
{\rm 1~trilinear~coupling}&: \; A \, ,  \\
{\rm Higgs~mixing~parameter}&: \; \mu \, ,  \nonumber \\
{\rm pseudoscalar~Higgs~mass}&: \; \MA \, ,  \nonumber \\
{\rm ratio~of~vevs}&: \; \tb \, ,  \nonumber
\end{align}
where $q_{1,2} \equiv u, d, s, c$, we assume soft SUSY-breaking
parameters for left- and right-handed sfermions, 
and the sneutrinos have the same soft SUSY-breaking parameter as the
corresponding charged sfermions. All of these parameters are specified
at a renormalisation scale $\msusy$ given by
the geometric mean of the masses of the scalar top eigenstates,
$\msusy \equiv \sqrt{\mst1 \mst2}$, which is also the scale at which
electroweak symmetry breaking conditions are imposed. 
The sampled parameter ranges are shown in \refta{tab:pmssm11-ranges}.

\begin{table}[thb!]
  \begin{center}
    \renewcommand{\arraystretch}{1.25}
  \begin{tabular}{|c|c|c|} \hline
Parameter   &  \; \, Range      & Segments \\
\hline
$M_1$       &  $(-4 ,  4 )\tev$  & 6 \\
$M_2$       &  $( 0 ,  4 )\tev$  & 2 \\
$M_3$       &  $(-4 ,  4 )\tev$  & 4 \\
$\msq{}$      &  $( 0 ,  4 )\tev$  & 2 \\
$\msqt{}$     &  $( 0 ,  4 )\tev$  & 2 \\
$\mslep{}$    &  $( 0 ,  2 )\tev$  & 1 \\
$m_{\tilde \tau}$       &  $( 0 ,  2 )\tev$  & 1 \\
$\MA$       &  $( 0 ,  4 )\tev$  & 2 \\
$A$         &  $(-5  , 5 )\tev$  & 1 \\
$\mu$        &  $(0 , 5 )\tev$  & 1 \\
$\tb$         &  ( 1  , 60)      & 1 \\
\hline \hline
\multicolumn{2}{|c|}{Total number of boxes} & 384     \\
\hline
  \end{tabular}
\end{center}
\caption{\it The ranges of the pMSSM11 parameters sampled, together with
  the numbers of segments into which each range was divided, and the
  corresponding total number of sample boxes.} 
\label{tab:pmssm11-ranges}
\end{table}


\section{The {\tt MasterCode}}
\label{sec:mc}

We perform a global likelihood analysis of the various MSSM incarnations
including constraints from direct searches for SUSY particles at the
LHC, measurements of the Higgs boson mass and signal strengths, LHC
and LEP searches for additional SUSY Higgs bosons, precision electroweak
observables (including \gmt)
flavor constraints from $B$- and $K$-physics observables, the
cosmological constraint on the overall CDM density,
and upper limits on spin-independent and -dependent LSP-nuclear
scattering. We treat $m_t$, $\alpha_s$ and $M_Z$
as nuisance parameters. Details about the explicit constraints employed
in each of the analyses can be found in \citeres{mc13,mc14,mc16,mc15}.

The observables contributing to the likelihood are calculated using the
{\tt MasterCode} tool~\cite{mcold,mc9,mc10,mc11,mc12,mc13,mc14,mc15,mc16,mcweb},
which interfaces and combines consistently various public and private codes
using the SUSY Les Houches Accord (SLHA)~\cite{SLHA}. The following
codes are used in the analysis (for the specific versions employed in the
various analyses, see \citeres{mc13,mc14,mc16,mc15}): 
{\tt SoftSusy}~\cite{Allanach:2001kg}
for the spectrum, {\tt FeynWZ}~\cite{Svenetal} for the electroweak
precision observables,
{\tt FeynHiggs}~\cite{FeynHiggs}
for the Higgs sector and \gmt, {\tt SuFla}~\cite{SuFla} and
{\tt SuperIso}~\cite{SuperIso}
for the flavor physics observables, {\tt Micromegas}~\cite{MicroMegas}
for the CDM relic density, {\tt SSARD}~\cite{SSARD} for the
spin-independent and -dependent elastic scattering cross-sections 
\ssi\ and \ssd. The uncertainties in the cross-sections are derived
from a straightforward propagation of errors in in the input quantities
which determine the cross-section. The dominant 
uncertainties are  discussed in \citere{mc15}. 
{\tt SDECAY}~\cite{Sdecay} is used for calculating sparticle branching ratios,
and {\tt HiggsSignals}~\cite{HiggsSignals} 
and {\tt HiggsBounds}~\cite{HiggsBounds} for calculating
constraints on the SUSY Higgs sector. 
For the LHC searches we mainly rely on the {\tt Fastlim}
\cite{Papucci:2014rja} approach (see \citeres{mc13,mc14,mc16,mc15} for
details concerning each analysis).

The sampling is done using the {\tt MultiNest} package~\cite{multinest}.
The segments defined in \reftas{tab:mamsb-ranges}-\ref{tab:pmssm11-ranges}
define boxes in the respective parameter spaces, which are 
sampled. In order to ensure a smooth overlap between boxes
and eliminate features associated with their boundaries, we choose for
each box a prior such that 80\% of the sample has a flat distribution 
within the nominal box, and 20\% of the sample is in
normally-distributed tails extending outside the box.


\section{Predictions for the ILC and CLIC}
\label{sec:pred}

In this section we review the predictions for various SUSY and heavy
Higgs-boson masses and their implications for the ILC and CLIC.
The ILC is now proposed with a staged machine design, with the first
stage at $\sqrt{s} = 250 \gev$~\cite{ILC250}. However, here we will
focus on later stages with $\sqrt{s} = 500 \gev$ (ILC500) and a hypothetical
final stage with $\sqrt{s} = 1000 \gev$ (ILC1000). For CLIC we assume a
center-of-mass energy of $\sqrt{s} = 3 \tev$. 
In the plots in the subsections below several horizontal lines indicate
the reach of the future $e^+e^-$ colliders: The green lines shows the reach
for SUSY particle pair production at the ILC500; the red line
corresponds to the ILC1000, the purple line to CLIC.
It should be kept in mind that by the production of a lighter and a
heavier SUSY particle (e.g.\ $e^+e^- \to \neu1\neu3$) the actual reach
can be higher than indicated by the horizontal lines.


\subsection{Predictions for the mAMSB}
\label{sec:pred-mamsb}

As discussed in \refse{sec:mAMSB}, within this mAMSB framework one finds 
$\mneu1 \simeq 3 \tev$ for wino DM and $\mneu1 \simeq 1.1 \tev$ for
Higgsino DM. Both yield very heavy mass spectra, where in the Higgsino
DM case the $\neu1$ and $\cha1$ are in the (pair production) reach of
CLIC. 
If, on the other hand, the LSP is not the only component of the CDM,
$\mneu1$ may be smaller, and $m_{3/2}$ may also be lowered 
substantially, and the overall mass spectrum can be substantially
lighter. We concentrate on this more favorable case here.

\begin{figure}[htb!]
\begin{center}
\includegraphics[width=1.1\textwidth,height=6.4cm]{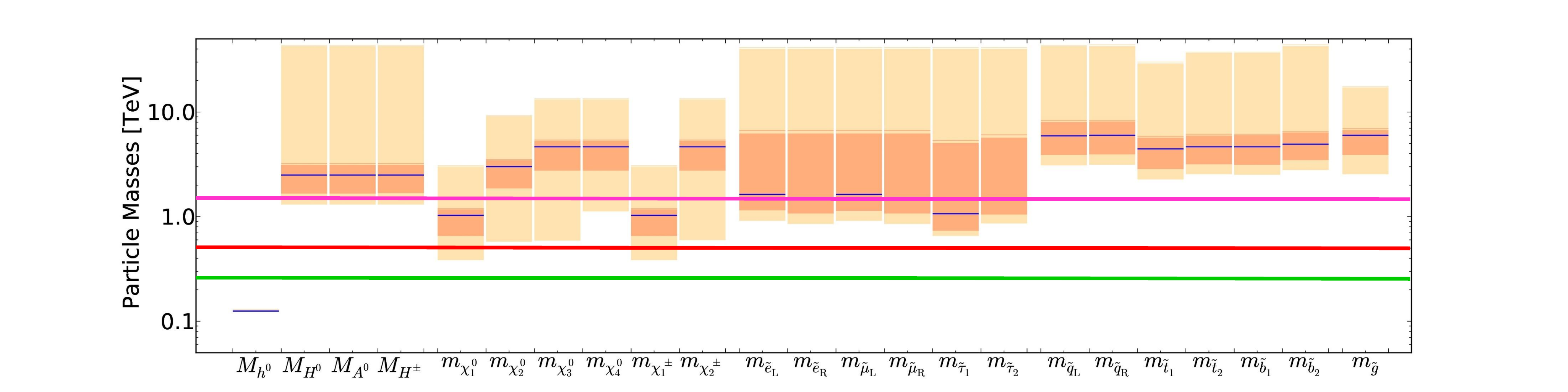}\\[1em]
\includegraphics[width=1.1\textwidth,height=6.4cm]{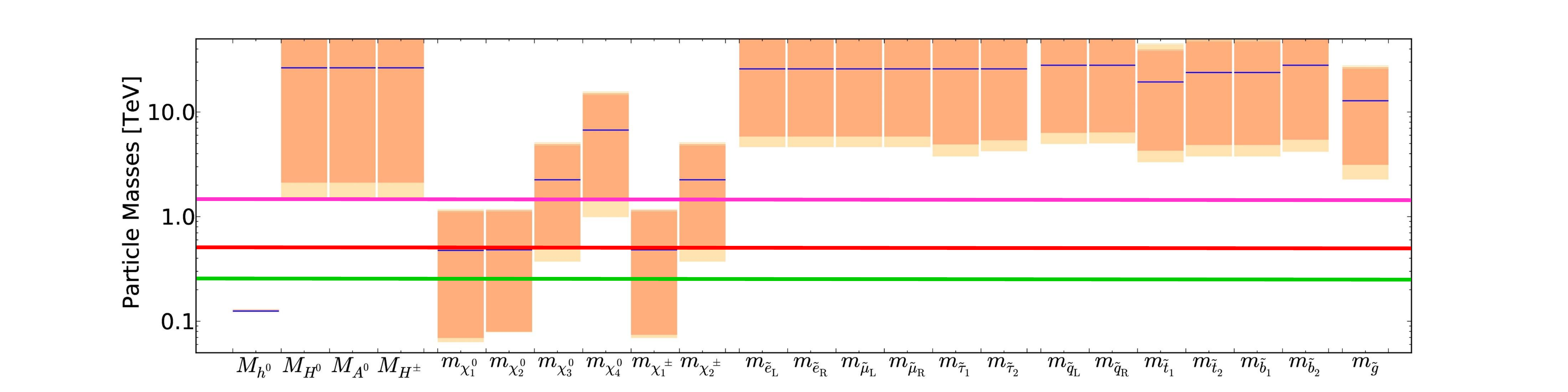}
\caption{Best-fit spectra (taken from \citere{mc13}) 
  in the mAMSB for wino DM (upper plot) and
  Higgsino DM (lower plot), relaxing the assumption that the LSP
  contributes all the CDM density. The one- and two-$\sig$~CL regions
  are shown in dark and light orange respectively, and the best-fit
  values are represented by horizontal blue lines. The green lines shows
  the reach for SUSY particle pair production at the ILC500; the red line
  corresponds to the ILC1000, the purple line to CLIC.
}
\label{fig:mAMSB}
\end{center}
\end{figure}

The overall fit using {\tt MasterCode} yielded for both DM cases a
$\chi^2/$dof of about 36.5/27, corresponding to a $p$-value of about
$0.105$~\cite{mc13}. 
In \reffi{fig:mAMSB}~\cite{mc13} we show the best-fit spectra in the
mAMSB for wino DM (upper plot) and 
Higgsino DM (lower plot), relaxing the assumption that the LSP
contributes all the CDM density. The one- and two-$\sig$~CL regions
are shown in dark and light orange respectively, and the best-fit
values are represented by blue lines. 
The colored horizontal lines indicate the reach of the future $e^+e^-$
colliders, as described above.

In the wino DM case one can see that the ILC has hardly any chance to see SUSY
particles, and might observe the $\neu1$, $\cha1$ and $\stau1$, with the
other sleptons having part of the $1\,\sig$ range within the kinematical
reach. However, all SUSY particles can be outside the reach even of CLIC
at the $2\,\sig$~level.
In the Higgsino DM case the $\neu1$, $\neu2$ and $\cha1$ are nearly mass
degenerate and possibly in the reach of the ILC, and for sure in the
reach of CLIC, which could also possibly observe the $\neu3$ and
$\cha2$. All other SUSY particles and heavy Higgs bosons are likely
outside the reach of CLIC. 
Overall the reach for SUSY particles at future $e^+e^-$
colliders is not overly favorable in the mAMSB.

\subsection{Predictions for the GUT based SU(5)}
\label{sec:pred-su5}

\begin{figure}[htb!]
\begin{center}
\includegraphics[width=1.1\textwidth,height=6.5cm]{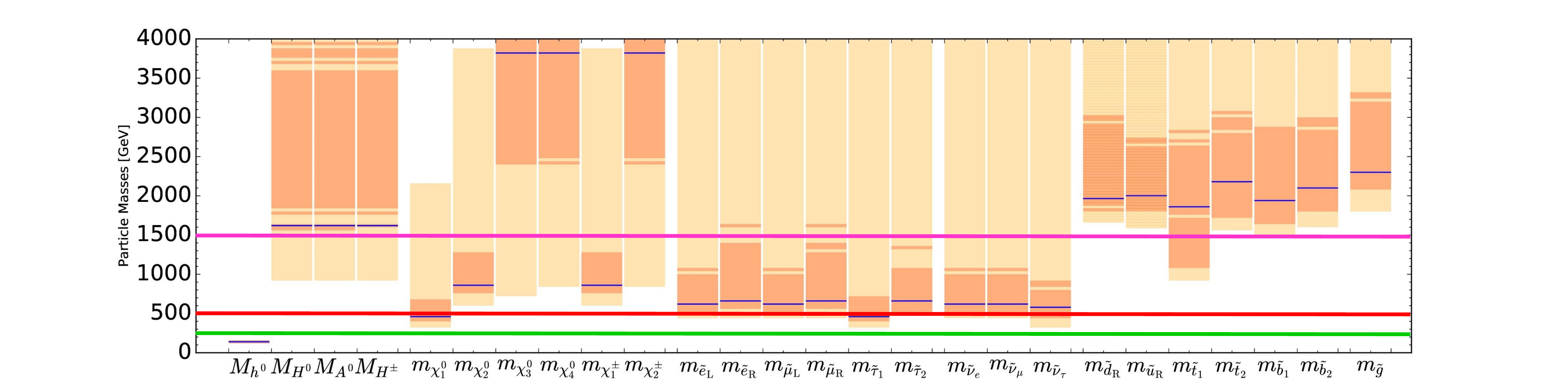}
\caption{Best-fit spectrum (taken from \citere{mc14}) 
  in the GUT based SU(5) model. The color coding is as in the previous plot.
}
\label{fig:SU5}
\end{center}
\end{figure}

The overall fit in the SU(5) model using {\tt MasterCode} yielded 
a $\chi^2/$dof of about 32.4/23, corresponding to a $p$-value of about
$0.09$~\cite{mc14}. The mass predictions for the GUT based SU(5) model
are shown in \reffi{fig:SU5}~\cite{mc14} with the color coding is as in
\reffi{fig:mAMSB}. As before the green/red/purple lines shows the reach
for SUSY particle pair production at the ILC500/ILC1000/CLIC.
Contrary to the mAMSB case we now demand that the LSP saturates the
Planck CDM bound. One can see that the ILC500 cannot produce any new
SUSY particles, while the ILC1000 might have the $\neu1$ and $\stau1$ in
reach. CLIC, on the other hand, with its higher center-of-mass energy
could possibly observe all sleptons (which masses are correlated), as
well as the $\neu{1,2}$ and $\cha1$. However all SUSY particles could be
out of reach even of CLIC at the $2\,\sig$~level.

\subsection{Predictions for the sub-GUT model}
\label{sec:pred-subGUT}

\begin{figure}[htb!]
\begin{center}
\includegraphics[width=1.1\textwidth,height=6.5cm]{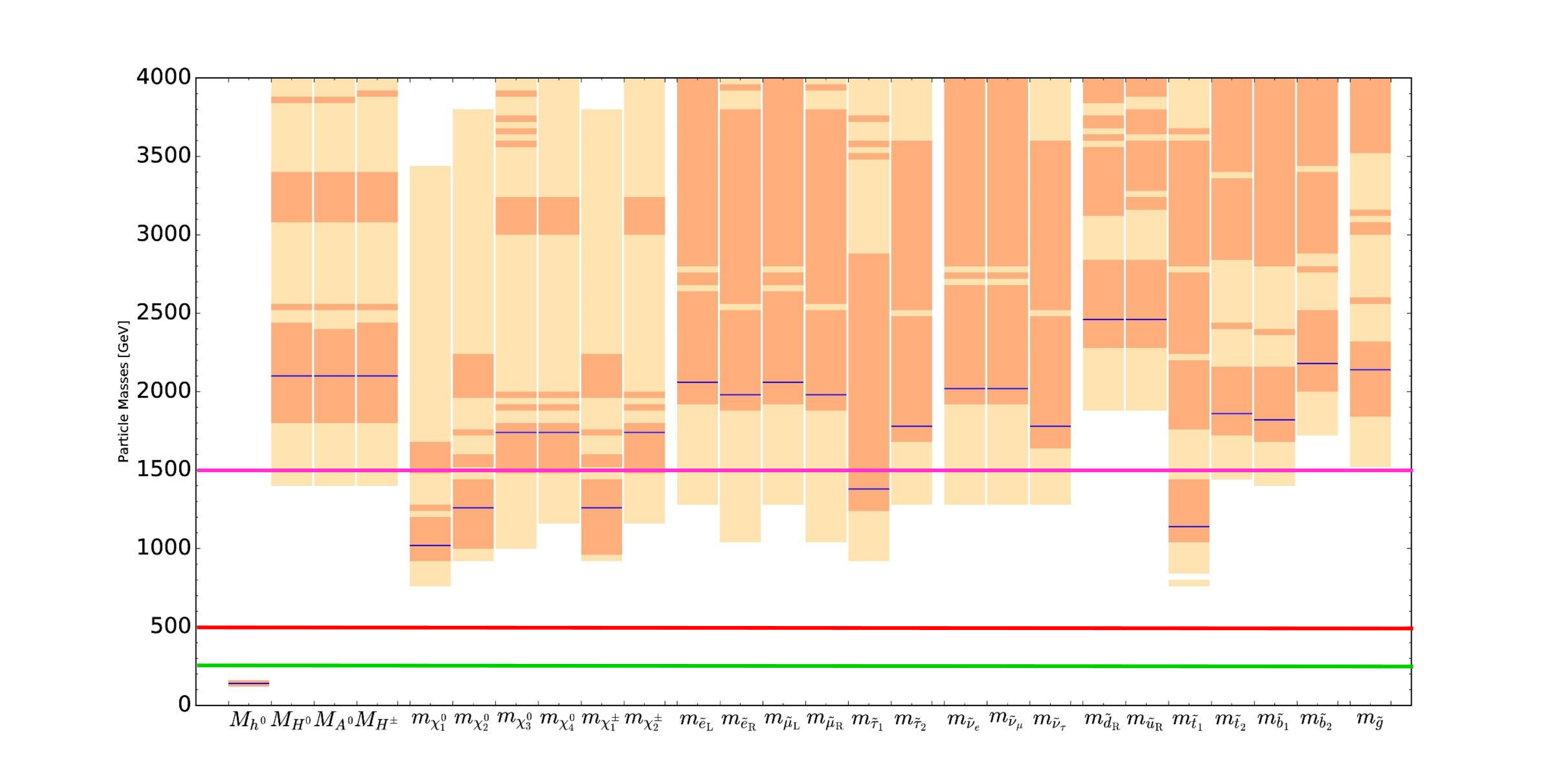}
\caption{Best-fit spectrum (taken from \citere{mc16}) 
  in the sub-GUT model. The color coding is as in the previous plots.
}
\label{fig:subGUT}
\end{center}
\end{figure}

The overall fit in the sub-GUT model using {\tt MasterCode} yielded 
a $\chi^2/$dof of about 28.9/24, corresponding to a $p$-value of about
$0.23$~\cite{mc16}. Part of the reason for the better value as compared
to the other two GUT based models is that the best-fit point of the
sub-GUT model is in somewhat better agreement with the measurement of 
\bmm, where the experimental value is slightly below the SM prediction.
In \reffi{fig:subGUT}~\cite{mc16} the 
{\tt MasterCode} results for the mass predictions in the sub-GUT model 
are displayed, with the color coding as in the previous
subsections. Also in this model one can observe that the ILC will not
have sufficient energy to produce SUSY particles. CLIC covers parts of
the $1\,\sig$~ranges of the $\neu{1,2}$, $\cha1$, $\stau1$ and
$\stop1$. However, all new particles have $1\,\sig$~ranges outside the
reach of CLIC. 
The overall reach for SUSY particles at future $e^+e^-$
colliders is not veryy favorable in the sub-GUT model.

\subsection{Predictions for the pMSSM11}
\label{sec:pred-pmssm11}

\begin{figure}[htb!]
\begin{center}
\includegraphics[width=1.0\textwidth,height=6.5cm]{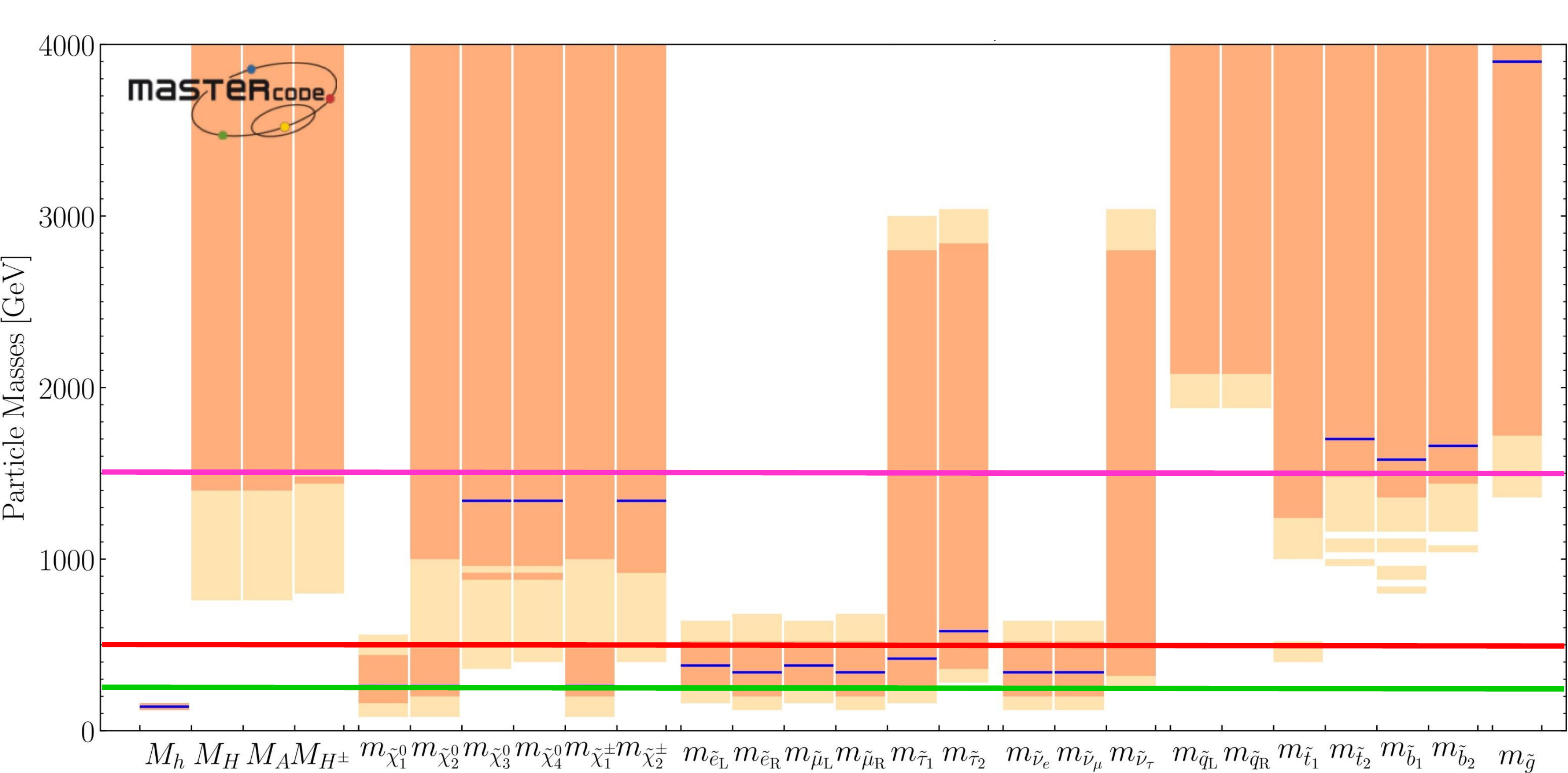}
\caption{Best-fit spectrum (taken from \citere{mc15}) 
  in the pMSSM11. The color coding is as in the previous plots.
}
\label{fig:pMSSM11}
\end{center}
\end{figure}

Finally we turn to the low energy model with 11 independent parameters,
the pMSSM11. 
The overall fit in the pMSSM11 using {\tt MasterCode} yielded 
a $\chi^2/$dof of about 22.1/20, corresponding to a $p$-value of about
$0.33$~\cite{mc15}. The mass predictions of the {\tt MasterCode} fit in
the pMSSM11 are shown in \reffi{fig:pMSSM11}~\cite{mc15}, with the color
coding as in the previous subsections. 
Contrary to the GUT based models in the pMSSM11 one can observe that
even the ILC500 has the chance to observe the $\neu1$, $\neu2$ and
$\cha1$. The ILC1000 covers the $1\,\sig$~ranges of the $\neu1$ and of
the sleptons of the first and second generation, and part of the
$1\,\sig$~ranges of the third generation sleptons. CLIC extends the
reach by covering part of the $1\,\sig$~ranges of the remaining
charginos and neutralinos, and touching the $1\,\sig$~ranges of some
squarks and the heavy Higgs bosons.

\section{Conclusions}
\label{sec:concl}

Using 
{\tt MasterCode}~\cite{mcold,mc9,mc10,mc11,mc12,mc13,mc14,mc15,mc16,mcweb}
we have performed global fits for various incarnations of the MSSM. 
{\tt MasterCode} combines consistently experimental data from LHC Higgs-boson
measurements, searches for additional Higgs bosons and supersymmetric (SUSY)
particles at the LHC, low-energy and flavor experiments as well 
as CDM measurements. 

Here we have reviewed the results of
studies of the GUT based mAMSB, of the GUT based SU(5) framework, 
of the sub-GUT model as well as of the pMSSM11, a phenomenological model
defined at low energies with 11 independent parameters. Details about
these studies can be found in \citeres{mc13,mc14,mc16,mc15}, respectively. 
In all models we have predicted the preferred SUSY
mass spectra at the $1$~and $2\,\sig$ level and compared them to the
reach of the ILC500, the ILC1000 and CLIC. 

We have found that the GUT based models offer substantially less
prospects to observe SUSY particles as compared to the pMSSM11. 
The ILC500 touches some $1\,\sig$~ranges in the GUT based models only in
the Higgsino DM case of the mAMSB, and the situation does not improve
substantially at the ILC1000, covering some best-fit points in the SU(5)
model and the Higgsino DM case of the mAMSB. CLIC has a higher reach,
but is mostly restricted to the lighter charginos and neutralinos in the
GUT based models, as well as the slepton spectrum in the SU(5) model.
In the pMSSM11, on the other hand, 
even the ILC500 could possibly observe the $\neu1$, $\neu2$ and
$\cha1$. The ILC1000 even covers the $1\,\sig$~ranges of the $\neu1$ and
of the first generation sleptons, as well as some parts of the 
$1\,\sig$~ranges of the sleptons of the third generation. CLIC also
covers part of the $1\,\sig$~ranges of the remaining
charginos and neutralinos, and have a reach even for some lighter 
squarks and the heavy Higgs bosons.

The {\tt MasterCode} analysis also yielded the $\chi^2/$dof values for
the various analyses. The resulting values are summarized in
\refta{tab:pvalue}. The three GUT based models have lowest $p$-values,
where the sub-GUT model performs somewhat better due to a better fit to
\bmm. Should this measurement go to the SM value, the $p$-value of the
sub-GUT model would become more similar to the ones of the mAMSB and the
SU(5) model. The pMSSM11 has the best $p$-value, in particular because
of a better fit of \gmt\ and the mass of the $W$~boson. In this model
the prospects of the potential future linear $e^+e^-$ colliders are
very good, with some electroweak particles even in the reach of the
ILC500. The pMSSM11, the model that fits the data best, would provide 
the possibility of many SUSY precision measurements. ``I~am very
optimistic.''~\cite{chance}

\begin{table}[t!]
\begin{center}
\begin{tabular}{|l|c|l|}
\hline
Model   &  $\chi^2/$dof   & $p$-value  \\
\hline 
mAMSB   & 36.5/27 & 0.105 \\
SU(5)   & 32.4/23 & 0.09  \\
sub-GUT & 28.9/24 & 0.23  \\
pMSSM11 & 22.1/20 & 0.33  \\
\hline
\end{tabular}
\caption{\it The $\chi^2/$dof values and the corresponding $p$-values
  of the four models under investigation.
}
\label{tab:pvalue}
\end{center}
\end{table}



\subsection*{Acknowledgements}

\begingroup 
We thank 
E.~Bagnaschi,
M.~Borsato, 
O.~Buchm\"uller, 
C.~Cavanaugh,
V.~Chobanova,
M.~Citron, 
J.C.~Costa, 
A.~De~Roeck,
M.~Dolan,
J.R.~Ellis,
H.~Fl\"acher,
G.~Isidori,
M.~Lucio,
F.~Luo,
D.~Mart\'inez-Santos,
K.A.~Olive,
A.~Richards, 
K.~Sakurai, 
V.~Spanos, 
I.~Su\'arez-Fern\'andez, 
K.J.~de~Vries
and 
G.~Weiglein
for the collaboration on the work presented here.
We thank M.~Mart\'inez for help with the plots.
The work of S.H.\ is supported 
in part by the MEINCOP Spain under contract FPA2016-78022-P, 
in part by the ``Spanish Agencia Estatal de Investigaci\'on'' (AEI) and the EU
``Fondo Europeo de Desarrollo Regional'' (FEDER) through the project
FPA2016-78022-P, 
and in part by
the AEI through the grant IFT Centro de Excelencia Severo Ochoa SEV-2016-0597.
During part of this work we used the middleware suite 
{\tt udocker}~\cite{udocker} to deploy {\tt MasterCode} on 
clusters, developed  by the EC H2020 project INDIGO-Datacloud (RIA 653549).
\endgroup



\end{document}